\documentclass[11pt]{article}

\usepackage{amssymb}

\newcommand{\be}{\begin{equation}}
\newcommand{\ee}{\end{equation}}
\newcommand{\lb}{\label}
\newcommand{\ol}{\overline}
\newcommand{\wt}{\widetilde}

\newcommand{\ba}{{\bf a}}
\newcommand{\bu}{{\bf u}}

\newcommand{\bx}{{\bf x}}
\newcommand{\br}{{\bf r}}

\newcommand{\bA}{{\bf A}}
\newcommand{\bB}{{\bf B}}

\newcommand{\bD}{{\bf D}}

\newcommand{\cM}{{\mathcal M}}

\newcommand{\bxi}{{\mbox{\boldmath $\xi$}}}

\newcommand{\grad}{{\mbox{\boldmath $\nabla$}}}
\newcommand{\bdot}{{\mbox{\boldmath $\cdot$}}}

\newcommand{\btimes}{{\mbox{\boldmath $\times$}}}
\newcommand{\bzed}{{\mbox{\boldmath $0$}}}

\newcommand{\scirc}{{\scriptstyle \circ}}
\newcommand{\sscirc}{{\scriptscriptstyle \circ}}

\begin{document}
\title{Turbulent Diffusion of Lines and Circulations}
 %\\ in the Kraichnan Velocity Ensemble}
\author{Gregory L. Eyink\\ Department of Applied Mathematics \& Statistics\\
The Johns Hopkins University\\
3400 N. Charles Street\\
Baltimore, MD 21218\\
Tel: 410-516-7201,\,\,\,Fax:  410-516-7459\\
e-mail: eyink@ams.jhu.edu}
\date{ }
\maketitle
\begin{abstract}
We study material lines and passive vectors in a model of turbulent flow at
infinite-Reynolds number, the  Kraichnan-Kazantsev  ensemble of velocities
that are white-noise in time and rough (H\"{o}lder continuous) in space.
It is argued that the phenomenon of ``spontaneous stochasticity'' generalizes
to material lines and that conservation of circulations  generalizes to a
``martingale property'' of the stochastic process of lines.


{\it keywords:} turbulence, material lines, circulations, Kraichnan model,
dynamo, fractals

\end{abstract}

\newpage

The evolution of material lines and surfaces passively carried by turbulent
flow has long been
a subject of interest \cite{Batchelor52}. This is motivated in part by
questions surrounding
dynamically relevant objects, such as vortex lines \cite{Taylor17,
TaylorGreen37, Taylor1938}
and magnetic field-lines \cite{Batchelor50, Saffman63}, which have been argued
to behave
similar to material lines at high Reynolds numbers.  However, in that limit,
the turbulent velocity
field is no longer differentiable in space but only H\"{o}lder continuous
\cite{Onsager49,
FrischParisi85, Eyink95a}. Observations from experiments and simulations
suggest that
material objects advected by such a rough velocity become fractal, with a
Hausdorff dimension
strictly greater than their topological dimension \cite{Mandelbrot76,
SreenivasanMeneveau86,
FungVassilicos91, VillermauxGagne94, Kivotides03, NicolleauElmaihy04}. This
poses a difficulty
to the view that vortex lines behave as material lines---a consequence of the
Kelvin-Helmholtz
theorem \cite{Helmholtz1858, Kelvin1869}---since circulations are {\it a
priori} not defined for
non-rectifiable loops. It has recently been argued that the Kelvin theorem in
fact breaks down
in turbulent flows, in the sense that the circulation is not strictly conserved
for every loop
\cite{Eyink06a, Chenetal06}. Similar breakdown of Alfv\'{e}n's theorem on
magnetic-flux
conservation \cite{Alfven42} is expected in plasma turbulence at high magnetic
Reynolds
numbers \cite{EyinkAluie06}.

These questions have been sharpened by recent work on the Kraichnan model of
advection
by a Gaussian random velocity field that is delta-correlated in time
\cite{Kraichnan68}. A novel
phenomenon has been discovered there called  {\it spontaneous stochasticity}:
Lagrangian particle
trajectories for a non-Lipschitz advecting velocity are non-unique and split to
form a random
process in path-space for a {\it fixed} velocity realization
\cite{Bernardetal98, GawedzkiVergassola00,
EVanden-Eijnden00, EVanden-Eijnden01, LeJanRaimond98, LeJanRaimond02,
LeJanRaimond04}.
This phenomenon raises many fundamental questions, including whether material
objects
such as lines and surfaces can even exist in the limit of infinite Reynolds
number. It is the purpose
of this Letter to outline a new approach to the evolution of such geometric
objects in the Kraichnan
model. We focus on material lines and passive vectors, which are dual objects
in the same sense
as material particles and passive scalar fields \cite{Falkovichetal01}. In
particular, we shall sketch
the proof of a ``martingale property'' that has previously been proposed
\cite{Eyink06a} as a
generalization of the conservation of circulations for a rough velocity field.

We consider stochastic flows \cite{Kunita90} on a $d$-dimensional manifold
$\cM$ driven by Brownian
vector fields that are not Lipschitz regular in space. To simplify the
presentation, we use Euclidean
space $\cM={\mathbb R}^d$ or the torus $\cM={\mathbb T}^d$ to illustrate the
main ideas. More precisely,
$\bu(\bx,t)$ is a Gaussian random vector field, with mean
$\overline{\bu}(\bx,t)$ and fluctuating part
$\widetilde{\bu}(\bx,t)$ with covariance
\be  \langle \widetilde{u}_i(\bx,t)\widetilde{u}_j(\bx',t')\rangle
                                                             =
D_{ij}(\bx,\bx';t)\delta(t-t'). \lb{u-cov} \ee
for $\bx,\bx'\in\cM.$ We are mainly interested in the case that
$\overline{\bu}(\bx,t)\equiv \bzed$
and $\bu(\bx,t)\equiv \widetilde{\bu}(\bx,t)$ is a homogeneous random field,
with $\bD(\bx,\bx';t)=
\bD(\bx-\bx',t).$ The quantity
\be \Delta(\bx,\bx';t)={\rm
tr}[\bD(\bx,\bx;t)+\bD(\bx',\bx';t)-2\bD(\bx,\bx';t)] \lb{Delta} \ee
is $\langle\|\bu(\bx)-\bu(\bx')\|^2\rangle,$ the mean of the Euclidean norm
squared, for a random field
$\bu(\bx)$ with covariance $\bD(\bx,\bx';t).$ The case of greatest interest to
us has $\Delta(\bx,\bx';t)
\propto \|\bx-\bx'\|^{2\alpha}$ with $0<\alpha<1.$ In that case, $\bu(\bx)$ is
H\"{o}lder
continuous with exponent $\alpha$ at every point in space.

We consider oriented lines (1-cells) given parametrically as continuous,
one-to-one maps
$C: \, [0,1]\rightarrow \cM.$  A {\it material line} satisfies
\be (d/dt)C(\sigma,t)=\bu( C(\sigma,t), \scirc \,t). \lb{line-eq} \ee
for $\sigma\in [0, 1]$ and $t\in {\mathbb R}.$ The circle ``$\scirc$'' means
that we interpret
equation (\ref{line-eq}) in the Stratonovich sense. The (forward) Ito equation
$(d/dt)C(\sigma,t)=
\bu(C(\sigma,t),t)$ equivalent to (\ref{line-eq}) has the mean changed to
$\overline{u}^*_i(\bx,t)=
\overline{u}_i(\bx,t)+(1/2)(\partial/\partial
x^k)D_{ik}(\bx,\bx';t)|_{\bx'=\bx}$ (\cite{Kunita90},
section 3.4.) If $\cM={\mathbb R}^d$ or ${\mathbb T}^d$ and if $\bu(\bx,t)$ is
a homogeneous random
field, then the Ito and Stratonovich interpretations of equation
(\ref{line-eq}) are equivalent.
Now let $P_\bu[C,t]$ denote the conditional probability distribution of lines
for a fixed velocity
realization $\bu.$ This distribution satisfies a {\it stochastic Liouville
equation}:
\be (d/dt)P_\bu[C,t]=-\int_0^1 d\sigma {{\delta}\over{\delta C_i(\sigma)}}
             \left(u_i(C(\sigma),\scirc\,t)\,P_\bu[C,t]\right).  \lb{Liouville}
\ee
Equation (\ref{Liouville}) is a direct consequence of equation (\ref{line-eq})
and must also be
interpreted in the Stratonovich sense. It is formally equivalent to the Ito
equation:
\begin{eqnarray}
(d/dt)P_\bu[C,t] &= &
-\int_0^1 d\sigma {{\delta}\over{\delta
C_i(\sigma)}}\left([\overline{u}^*_i(C(\sigma),t)
+\widetilde{u}_i(C(\sigma),t)]P_\bu[C,t]\right) \cr
& &\,\,\,\,\,\,\,\,\,\,\,\,\,\,+
{{1}\over{2}}\int_0^1 d\sigma \int_0^1 d\sigma' {{\delta^2}\over{\delta
C_i(\sigma)\delta C_j(\sigma')}}
\left(D_{ij}(C(\sigma),C(\sigma');t)P_\bu[C,t]\right). \lb{Ito-Liouville}
\end{eqnarray}
Averaging equation (\ref{Ito-Liouville}) over the Gaussian ensemble of
velocities $\widetilde{\bu}$
yields a functional {\it Fokker-Planck equation} for distributions in the space
of free lines
$C$ on the manifold $\cM:$
\begin{eqnarray}
(d/dt)P[C,t] &= &
-\int_0^1 d\sigma {{\delta}\over{\delta
C_i(\sigma)}}\left([\overline{u}^*_i(C(\sigma),t)P[C,t]\right) \cr
& &\,\,\,\,\,\,\,\,\,\,\,\,\,\,+
{{1}\over{2}}\int_0^1 d\sigma \int_0^1 d\sigma' {{\delta^2}\over{\delta
C_i(\sigma)\delta C_j(\sigma')}}
\left(D_{ij}(C(\sigma),C(\sigma');t)P[C,t]\right). \lb{FP-eq}
\end{eqnarray}
The first term on the righthand side represents a drift with the mean velocity
$\ol{\bu}^*$
and the second term represents a diffusion arising from the velocity covariance
$\bD.$
Similar diffusions on the path- and loop-spaces of a manifold $\cM$ have been
much studied,
motivated in part by questions from quantum field theory \cite{DriverRockner92,
Leandre02}.

The above considerations are rigorously justifiable for the case of a Lipschitz
velocity with
$\alpha=1$ but are only formal when $\alpha<1.$ A more careful (and also more
physically
realistic) approach in the latter case is to replace the advecting velocity
$\bu$ with a
``coarse-grained'' or smoothed velocity $\bu_\lambda =  \varphi_\lambda*\bu,$
by convolution
with a smooth filter kernel
$\varphi_\lambda(\br)=\lambda^{-d}\varphi(\br/\lambda).$ The length-scale
$\lambda$ can be interpreted as a mathematical representation of the viscous
cutoff in a true
turbulent velocity field \cite{EVanden-Eijnden00, EVanden-Eijnden01}. The exact
solution
of the Liouville equation (\ref{Liouville}) for such a smoothed velocity is
\be P_{\bu_\lambda} (dC,t | C_0, t_0) =
\delta(C-\bxi^{t,t_0}_{\lambda}(C_0))dC,
\lb{delta-sol} \ee
with initial condition $P_{\bu_\lambda} (dC,t_0 | C_0, t_0) = \delta(C-C_0)dC.$
Here
$\bxi^{t,t'}_{\lambda}:\cM\rightarrow \cM$ is the stochastic flow of
diffeomorphisms
generated by the smoothed velocity-field $\bu_\lambda$ (\cite{Kunita90},
section 4.6).
Despite (\ref{delta-sol}), a nontrivial diffusion process in line-space can be
obtained
if the limit $\lambda\rightarrow 0$ is taken appropriately. Consider a ``nice''
distribution
$G_\rho(dC)$ which is supported on lines entirely contained in the ball
$B(\bzed,\rho)$
of radius $\rho$ at the origin $\bzed$ and take the weak limit
\be \lim_{\rho\rightarrow 0}\lim_{\lambda\rightarrow 0}
      \int G_\rho (dC'_0) \int P_{\bu_\lambda} (dC,t | C_0+C_0', t_0) \Psi(C)
      = \int P_\bu(dC,t | C_0, t_0) \Psi(C) \lb{weak-lim} \ee
for bounded, continuous functionals $\Psi(C)$ and $t>t_0.$ In the ``weakly
compressible
regime'' \cite{GawedzkiVergassola00, EVanden-Eijnden00, LeJanRaimond02}---
and, in particular, for a divergence-free velocity field satisfying
$\grad\bdot\bu=0$---
this limit should yield a non-degenerate diffusion. \footnote{We note that the
same
diffusion process should also be obtained by using Duhamel's formula to solve
equation (\ref{Ito-Liouville}) as the Ito integral
\be
P_\bu[C,t] = S_0^*(t,t_0) P[C,t_0] - \int_{t_0}^t dt'\,S^*(t,t')
    \int_0^1 d\sigma {{\delta}\over{\delta
C_i(\sigma)}}\left([\widetilde{u}_i(C(\sigma),t')]
                                P_\bu[C,t']\right),  \lb{Duhamel} \ee
where  $S_0^*(t,t')={\rm Texp}\left[\int_{t_0}^t dt'\,{\cal A}^*(t')\right]$
and
\be  {\cal A}^*(t)= -\int_0^1 d\sigma {{\delta}\over{\delta
C_i(\sigma)}}\left([\overline{u}^*_i(C(\sigma),t)\cdot\right)
+{{1}\over{2}}\int_0^1 d\sigma \int_0^1 d\sigma' {{\delta^2}\over{\delta
C_i(\sigma)\delta C_j(\sigma')}}
\left(D_{ij}(C(\sigma),C(\sigma');t)\cdot\right) \ee
is the Fokker-Planck operator of the (mean) diffusion in line-space.
Equation (\ref{Duhamel}) can be solved iteratively to generate a representation
$P_\bu[C,t]=S_\bu^*(t,t_0)P[C,t_0]$ as a Wiener chaos expansion in white-noise
$\widetilde{\bu};$
cf. \cite{LeJanRaimond02,LeJanRaimond04}.} This is a generalization of the
phenomenon of  spontaneous stochasticity to the turbulent advection of lines,
with
initial line $C_0$ at time $t_0$ splitting into a random ensemble of lines $C$
at time $t.$

As for the case of smooth advection, an unconditional diffusion satisfying
equation (\ref{FP-eq})
may be obtained by averaging over the velocity $\bu.$  The instantaneous
realizations $C$
of this diffusion process should be fractal objects when the advecting velocity
is H\"{o}lder
continuous with exponent $0<\alpha<1$ and rigorous estimates of their Hausdorff
dimensions
would be of much interest. These questions may also be addressed numerically
using
Lagrangian Monte Carlo techniques \cite{Frischetal98, Frischetal99,Gatetal98}.
In such
a study, the material line $C(t)$ would be represented by a discrete
approximation $C_{N}(t)$
constructed from $N+1$ Lagrangian particles $\bx_a(t),\,\,a=0,...,N:$
\be C_{N}(\sigma,t) = (1-\theta_{N}(\sigma))\bx_{a_{N}(\sigma)}(t)
      + \theta_{N}(\sigma)\bx_{a_{N}(\sigma)+1}(t). \lb{discrete} \ee
Here $a_{N}(\sigma)=[N\sigma]$ with $[x]$ the greatest integer less than or
equal to $x$ (modulo $N$ for loops) and  $\theta_{N}(\sigma)=(N\sigma)$
where $(x)=x-[x]$ is the fractional part of $x.$ Thus, (\ref{discrete})
corresponds to
a piecewise-linear curve with linear segments connecting the successive
Lagrangian
particles. So long as $\delta_N(t)=\max_a
|\bx_a(t)-\bx_{a+1}(t)|\lesssim\lambda,$
the discrete approximation $C_{N}(t)$ represents well the material line $C(t)$
and the
approximation becomes better as $N\rightarrow\infty$ and $\delta_N(t)
\ll \lambda$. However, the same is not true in the opposite limit, where
$\lambda\ll \delta_N(t).$ The phenomenon of  ``spontaneous stochasticity'' for
a
rough velocity field makes it very doubtful that material lines even exist  if
the
limit $\lambda\rightarrow 0$ is taken before evolving in time and an initial
line
then presumably ``explodes'' into a disconnected cloud of particles at any time
$t>0.$ Thus, the velocity smoothing in (\ref{weak-lim}) appears to be necessary
to define appropriately a line-diffusion for a rough (H\"older) velocity.
Alternatively,
a stochastic regularization might be employed that adds a white-noise $\kappa\,
dW(t)$ to the evolution equation of Lagrangian particles
\cite{Falkovichetal01}.

We now turn to the dual problem of a passive vector (1-form) $\bA$ advected by
the
velocity $\bu=\ol{\bu}+\widetilde{\bu}$:
\be \partial_t \bA(\bx,t)+ (\bu(\bx,\sscirc t)\bdot\grad)\bA(\bx,t)+
               (\grad\bu(\bx,\sscirc t))\bA(\bx,t)=\bzed, \lb{vec-eq} \ee
This stochastic equation is interpreted again in the Stratonovich sense.
Equation (\ref{vec-eq})
for $d=3$ is equivalent by vector calculus identities to $\partial_t \bA
+\grad(\bu\bdot\bA)
-\bu\btimes(\grad\btimes\bA)=\bzed.$ The latter has the form of Ohm's law,
\be {\bf E}+\bu\btimes \bB = \eta {\bf J}, \lb{ohm} \ee
in the ideal limit of zero resistivity $(\eta=0)$ for an electric field ${\bf
E}= -\partial_t \bA
-\grad(\bu\bdot\bA)$ and magnetic field $\bB=\grad\btimes\bA$ given by a vector
potential $\bA,$
with the electric current ${\bf J}=\grad\btimes\bB.$
Taking the curl of (\ref{ohm}) yields an induction equation $\partial_t \bB
=\grad\btimes
(\bu\btimes\bB) + \eta \bigtriangleup \bB$ for the magnetic field. With this
interpretation,
the passive vector equation was introduced by Kazantsev \cite{Kazantsev68} as a
soluble
model of the kinematic dynamo. (See also \cite{Vergassola96, Vincenzi02,
Celanietal06}.)
The ``circulation'' (or  ``holonomy") of $\bA$ along $C$ is defined in
Lagrangian form as
\be \Phi_L(C,t)=\int_{C(t)}\,\bA(t)\bdot d\bx,  \lb{B-flux} \ee
where $C(t)$ is the material line advected by $\bu(t)$ which started as line
$C$
at the initial time $t=t_0.$ Conservation of  ``circulation",
$(d/dt)\Phi_L(C,t)=0,$ follows formally
from (\ref{vec-eq}) for any space dimension $d\geq 1.$  It is rigorously true
for the case of a
smooth advecting velocity with $\alpha=1$ (\cite{Kunita90}, section 4.9). If
$C$ is a closed loop
(1-cycle), then the line-integral (\ref{B-flux}) represents gauge-invariant
magnetic flux and
the conservation law corresponds to Alfv\'{e}n's theorem \cite{Alfven42}.

We now consider the generalization of this result for $\alpha<1.$ For this
purpose,
it is useful to reformulate the passive vector equation (\ref{vec-eq}) as a
{\it passive
scalar in line-space}:
\be  \partial_t \Phi_E(C,t)+\int_0^1 d\sigma
\,\,[\ol{u}_i(C(\sigma),t)+\wt{u}_i(C(\sigma),\scirc t)]
       \frac{\delta}{\delta C^i(\sigma)}\Phi_E(C,t) =0. \lb{line-scalar} \ee
In this equation, $\Phi_E(C,t)=\int_C \bA(t)\bdot d\bx$ is the Eulerian
circulation of $\bA$ along
a {\it fixed} (non-advected) line $C.$ An exactly analogous reformulation of
the incompressible
Euler equation (as an active scalar in loop-space) was advanced some time ago
by Migdal
\cite{Migdal93}.  Note that conservation of circulations is just the formal
solution of (\ref{line-scalar})
by the method of characteristics.  We shall take the equation
(\ref{line-scalar}) as our primitive
formulation of the passive vector; one of the immediate advantages is that we
can avoid
(for the moment) the question how to define line-integrals over fractal lines.
We then
convert (\ref{line-scalar}) to Ito formulation:
\begin{eqnarray}
\partial_t \Phi_E(C,t) & = & -\int_0^1 d\sigma
\,\,[\ol{u}_i^*(C(\sigma),t)+\wt{u}_i(C(\sigma),t)]
                                       \frac{\delta}{\delta
C^i(\sigma)}\Phi_E(C,t)  \cr
        & & \,\,\,\,\,\,\,\,\,\,\,
        +\frac{1}{2}\int_0^1 d\sigma \int_0^1 d\sigma' \,\,
       D_{ij}(C(\sigma),C(\sigma'),t) \frac{\delta^2}{\delta C^i(\sigma)\delta
C^j(\sigma')}\Phi_E(C,t)
\end{eqnarray}
This stochastic equation is solved by the method of LeJan and Raimond
\cite{LeJanRaimond02,
LeJanRaimond04} (cf. footnote \#1), writing it as a (backward) Ito integral and
iterating  to obtain
$\Phi_E(C,t) = S_\bu(t,t')\Phi_E(C,t'),$ where the Markov operator semi-group
$S_\bu(t,t')$ is
defined by a Wiener chaos expansion. More  intuitively, this solution is
expressed as
\be \Phi_E(C,t)=\int P_\bu(dC',t' | C,t) \Phi_E(C',t'),\,\,\,\,\,\,\,\, t'<t,
\lb{stochastic-Alfven} \ee
in terms of the turbulent diffusion of lines (backward in time). We see that
the circulations
are {\it not} conserved, except on average. This is precisely the ``martingale
property'' that
was conjectured (for solutions of incompressible Euler equations) in
\cite{Eyink06a}. Note
that this property imposes an irreversible arrow of time, since Eulerian
circulations are
given as averages over their past values, not future ones. This ``generalized
Alfv\'{e}n
theorem'' should be related to dynamo action in the Kazantsev model
\cite{Kazantsev68,
Vergassola96, Vincenzi02, Celanietal06}. In the physical context of the dynamo,
there is
an additional resistive term $\eta\oint_{C} dx_j \partial_i
\frac{\delta}{\delta\sigma_{ij}(\bx)}
\Phi_E(C,t)$ on the righthand side of (\ref{line-scalar}), where
$\delta/\delta\sigma_{ij}(\bx)$
is the ``area derivative'' in the loop calculus of Migdal \cite{Migdal93}. This
term breaks
time-reversal symmetry and should select the backward-martingale solution
(\ref{stochastic-Alfven}) in the ideal limit $\eta\rightarrow 0.$ Of course, it
cannot
be ruled out {\it a priori} that the $\eta\rightarrow 0$ limit of the resistive
regularization
and the $\lambda\rightarrow 0$ limit for regularized velocity, as in
(\ref{weak-lim}),
shall yield distinct weak solutions of the loop-equation (\ref{line-scalar}),
as occurs
in the intermediate compressibility regime of the passive scalar problem
\cite{GawedzkiVergassola00,EVanden-Eijnden00,EVanden-Eijnden01,LeJanRaimond02}.

These results can be generalized to turbulent diffusion processes of
higher-dimensional
material objects, $k$-dimensional oriented submanifolds of $\cM$ or {\it
$k$-cells} $C^k(t).$
The dual object is the {\it passive k-form} $\omega^k$, which satisfies (in
Stratonovich sense)
\be \partial_t\omega^k + L_\bu\omega^k =0 \lb{Lie-deriv} \ee
with $L_\bu$ the Lie-derivative along the vector field $\bu$ (\cite{Kunita90},
section 4.9).
This equation is formally equivalent to conservation of the integral invariants
\be I(C^k,t)=\int_{C^k(t)}\,\omega^k(t) \lb{k-invariant} \ee
for any $k$-cell $C^k(t)$ comoving with $\bu$ \cite{Abrahametal83}. Then $k=0$
is the passive
scalar, $k=1$ the passive vector and $k=d$ the passive density
\cite{GawedzkiVergassola00}.
A theory similar to that developed here for $k=1$ applies for any integer $k.$
A unified
approach to all these results is to consider directly the turbulent diffusion
of the Lagrangian
flow maps $\bxi^{t,t'},$ which satisfy the stochastic equation
\be (d/dt)\bxi^{t,t'}(\ba)=\bu(\bxi^{t,t'}(\ba),\sscirc
t),\,\,\,\,\,\,\,\,\bxi^{t',t'}(\ba)=\ba. \ee
In this framework one can derive for the distribution $P_\bu[\bxi,t]$  on maps
exact
analogues of the Liouville  equation, in Stratonovich form (\ref{Liouville}) or
Ito form
(\ref{Ito-Liouville}). It is natural to formulate the problem as an
infinite-dimensional diffusion in the
Hilbert space ${\cal H}=L^2(\cM,{\mathbb R}^d)$ . It is known for the cases
$\cM={\mathbb R}^d$
or ${\mathbb T}^d$ that  the semigroup $S(\cM)$ of Borel volume-preserving maps
is a
closed subset of this Hilbert space, and that the group $G(\cM)$ of
volume-preserving
diffeomorphisms is dense in $S(\cM)$ for the $L^2$-topology \cite{Brenier03}.
This
construction is a close analogue of the ``generalized Euler flows'' of
Brenier, but for the
Cauchy initial-value problem.

To summarize: We have outlined an approach to the study of material lines in a
model
of turbulent flow at infinite Reynolds number and to the dual problem of a
passive vector in
the same flow. The main conclusions are (1) that a non-degenerate diffusion
should exist
for material lines, generalizing the phenomenon of ``spontaneous
stochasticity'' of
material points, and (2) that the Kelvin/Alfv\'{e}n theorem on conservation of
circulations
should generalize to a ``martingale property''. Although the approach sketched
here
depends heavily on the white-noise character of the velocity field in time, we
expect
that similar results hold for more realistic velocity ensembles with the
crucial property
that realizations are rough (H\"{o}lder continuous) in space. See
\cite{Eyink06a,EyinkAluie06}
for related rigorous results on the solutions of incompressible fluid
equations. The two properties
discussed in the context of this model problem should be an essential feature
of real fluid
turbulence in the high Reynolds number limit.

\bigskip

\noindent {\bf Acknowledgements.} We thank S. Chen, M. Chertkov, L. Chevillard,
R. Ecke, C. Meneveau, K. R. Sreenivasan and E. T. Vishniac for useful
conversations.
This work was supported by the NSF grant \# ASE-0428325 at the Johns Hopkins
University and by the Center for Nonlinear Studies at Los Alamos National
Laboratory,
where the research was begun.

\end{document}